\documentclass[prl,twocolumn,preprintnumbers,amsmath,amssymb,showpacs,superscriptaddress]{revtex4} 
\usepackage{epsfig}
\usepackage{amssymb}
\usepackage{amsmath}
\usepackage{bm}
\usepackage{color}
\usepackage{times}
\usepackage[colorlinks,bookmarks=false,citecolor=blue,linkcolor=red,urlcolor=blue]{hyperref}
\newcommand{\be}{\begin{equation}}
\newcommand{\ee}{\end{equation}}
\newcommand{\bea}{\begin{eqnarray}}
\newcommand{\eea}{\end{eqnarray}}

\def\fr#1{(\ref{#1})}

\begin{document}
\title{Multi-particle bound state formation following a quantum quench
to the one-dimensional Bose gas with attractive interactions}
\author{Lorenzo Piroli}
\author{ Pasquale Calabrese}
\affiliation{SISSA and INFN, via Bonomea 265, 34136 Trieste, Italy. }
\author{Fabian H.L. Essler}
\affiliation{The Rudolf Peierls Centre for Theoretical Physics, Oxford
University, Oxford OX1 3NP, UK}

\begin{abstract}
We consider quantum quenches from an ideal Bose condensate to the
Lieb-Liniger model with arbitrary attractive interaction strength. We
focus on the properties of the stationary
state reached at late times after the quench. Using recently developed
methods based on integrability, we obtain an exact description of
the stationary state for a large number of bosons. A distinctive
feature of this state is the presence of a hierarchy of multi-particle
bound states. We determine the dependence of their densities on
interaction strength and obtain an exact expression for the stationary
value of the local pair correlation $g_2$. We discuss ramifications of
our results for cold atom experiments.
\end{abstract}
\maketitle

Integrable models have a long and venerable history of providing
crucial points of reference that have greatly facilitated our
understanding of interacting many-particle quantum systems.
While integrable models are by definition special, recent advances in
the field of ultra cold atoms have made it possible to realize them
to a good approximation experimentally
\cite{review,kinoshita,ki-05,YY-chip,fabbri,ising-exp,yg-exp}. Moreover,
small deviations from integrability normally only lead to small
changes in experimentally observable quantities.

One of the most striking features of quantum integrable models is that
they typically feature hierarchies of bound states that often involve
arbitrary numbers of elementary particles \cite{Tbook,Mbook}. Such
bound states have proved to be difficult to observe in e.g. neutron
scattering experiments on quantum spin chain materials, because
their signatures in equilibrium dynamics are often small.  
Recent experimental advances in cold atomic gases
have made it possible to observe non-equilibrium dynamics in isolated
many-particle quantum systems in exquisite detail
\cite{qnc,STG,cetal-12,getal-11,shr-12,mmk-13,fse-13, langen-13,agarwal-14,geiger-14, silva}. Integrable
systems have again played a key role in these developments.
In particular, following the theoretical proposal of
Ref.~\cite{gree12}, signatures of propagating two particle bound
states following a local quench in an (almost) integrable system were
successfully observed experimentally \cite{fse-13}. One of the most
exciting aspects of non-equilibrium dynamics in integrable systems is
that it allows one to realize new stable states of matter. One
remarkable example is the so called super Tonks-Girardeau gas,
obtained at late times after quenching a Bose gas from an infinite
repulsive interaction to an infinite attractive one
\cite{abcg-05,STG}. This new state of matter has truly remarkable
properties \cite{abcg-05,bbgo-05,mf-10,kmt-11,pdc-13,th-15},
exhibiting stronger correlations than the repulsive Tonks-Girardeau
gas (which has also been probed experimentally \cite{kinoshita}).
In this letter we investigate a protocol very similar to the super
Tonks-Giradeau case, i.e. a quench from non-interacting to attractive
Bose gas. We show that the stationary state reached at late times
after the quench is a novel state of matter distinguished by a
characteristic distribution of the densities of multi-particle bound
states.

We consider a one-dimensional gas of $N$ bosons with attractive
point-like interaction, i.e. the Lieb-Liniger 
model \cite{ll}  
\begin{equation}
H = -\frac{\hbar^2}{2m}\sum_{j=1}^N \frac{\partial^2}{\partial {x_j^2}} 
- 2\bar c \sum_{ i<j} \delta(x_i - x_j).
\label{hamiltonian}
\end{equation}
Here $\bar c > 0$ is the interaction strength and  $m$ the mass of the
particles (atoms). The former is related to the effective 1D 
scattering length $a_{1D}$, which can be tuned experimentally via
Feshbach resonances \cite{o-98}, by $\bar c=\hbar^2/m a_{1D}$.
We consider a system of length $L$ with periodic
boundary conditions, and ultimately are interested in the limit $L\to
\infty$ while keeping the density of the gas $D=N/L$ fixed. For later
convenience we define the dimensionless coupling constant $\gamma=\bar c/D$. 
In the following, we fix $\hbar=2m=1$. 
Quantum quenches in the Lieb-Liniger model have been widely 
investigated in the literature
\cite{gritsev,mossel,iyer,ks-13,poz-14,m-13,de_nardis_wouters,cd-14,kcc14,dc-14,dpc-15,vdberg,ga-15},
but have mainly focused on the repulsive regime.  
The Lieb-Liniger model is solvable by Bethe ansatz for any value of
$\bar{c}$ \cite{ll}. The eigenstates (called Bethe states)
$\Psi_{\{\lambda_j\}}$ are parametrized by sets of $N$ complex
momenta $\{\lambda_j\}_{j=1}^{N}$, which satisfy the quantization
conditions (``Bethe equations'') 
\begin{equation}
e^{-i\lambda_jL}=\prod_{k\neq j}^{N}\frac{\lambda_k-\lambda_j-i\bar{c}}{\lambda_k-\lambda_j+i\bar{c}}\ ,\quad j=1,\ldots, N\ .
\label{bethe_eq}
\end{equation}
Bethe states are given by superposition of plane waves \cite{ll}
$\Psi_{\{\lambda_j\}}(x_1,...x_N) = \sum_P A_P \prod_{j=1}^N e^{i
  \lambda_{P_\ell} x_\ell} $, where the sum is over all permutations
$P$ of the rapidities $\{\lambda_j\}$ and the amplitudes are
$A_P=\prod_{N \geq \ell > k \geq 1} (1+ \frac{i \bar c
  ~\text{sgn}(x_\ell - x_k))}{\lambda_{P_\ell} - \lambda_{P_k}})$. 
In the attractive regime, the solutions of (\ref{bethe_eq}) arrange
themselves into patterns in the complex rapidity plane consisting of
``strings'' \cite{m-65,Tbook}. A general solution with $N$ rapidities
will consist of $N_j$ strings of length $j$, where $N=\sum_{j}jN_j$. A
single $j$-string takes the form
\begin{equation}
\lambda^{j,a}_{\alpha}=\lambda_{\alpha}^{j}+\frac{i\overline{c}}{2}(j+1-2a)+i\delta^{j,a} ,\quad a=1,\ldots, j .
\label{string_definition}
\end{equation}
Here $\alpha$ and $a$ respectively label the string under
consideration and the individual rapidities within that string,
while $\delta^{j,a}$ are exponentially small deviations in the system
size $L$. Following standard practice we will ignore these deviations. The string centers $\lambda_{\alpha}^{j}$ are real numbers and
fulfil a generalized Pauli principle that imposes all string centers
to be different for a given solution of \fr{bethe_eq}.
It follows from the Bethe ansatz form of the wave function that string
solutions correspond to multi-particle bound states in the sense that
the wave function decays exponentially with respect to the distance
between any two particles in the bound state.
The energy and momentum of an $N$-particle Bethe state consisting of
$N_j$ strings of length $j$ are
\be
K=\sum_{(j,\alpha)} j \lambda^j_{\alpha}\ ,\quad
E = \sum_{(j,\alpha)} j (\lambda^j_{\alpha})^2 - \frac{\bar c^2}{12} j(j^2 - 1).
\label{momentum_energy}
\ee
Eq. \fr{momentum_energy} shows that $\lambda^j_{\alpha}$ is the contribution
of each particle in a $j$-particle bound state to the total momentum.
The ground-state has zero momentum  and consists of a single
$N$-string \cite{Tbook}. While the thermodynamic limit in thermal
equilibrium does not exist \cite{Tbook}, 
correlation functions can be calculated at zero density \cite{cc-07}. 
Crucially, in the quantum quench context of interest here, the
infinite volume limit at fixed particle density does exist. In this
limit macro-states can be described in complete analogy to the
standard finite-temperature formalism \cite{Tbook} by particle and hole
densities $\{\rho_n(\lambda)\}$, $\{\rho^h_{n}(\lambda)\}$. In
particular, $\rho_n(\lambda)$ gives the distribution of $n$-string
centers of a macro-state which, in the thermodynamic limit, form a
dense set on the real line. Similarly, $\rho^h_n(\lambda)$ is the
distribution of holes of $n$-string centers. The latter is
analogous to the distribution of holes (i.e. unoccupied states) known
from the ideal Fermi gas at finite temperature. While in the case of
free fermions it is trivially related to the Fermi-Dirac distribution,
in the interacting case we are considering here, the relation between
$\rho_n(\lambda)$ and $\rho_n^h(\lambda)$ is more involved and derives
from the Bethe equations (\ref{bethe_eq}) \cite{korepin}. 

{\it The quench protocol}.
Our initial state is the ground state in the absence of interactions, i.e. the BEC state \cite{ks-13,de_nardis_wouters}. 
The evolution for $t>0$ is governed by the attractive Lieb-Liniger Hamiltonian. At time $t$, the expectation value of any observable $O$ is given by (denoting energy eigenstates with $|\mu\rangle$, $|\nu\rangle$)
\begin{equation}
\langle \Psi(t) | O| \Psi(t) \rangle= \sum_{\mu, \nu} \langle \Psi_0|\mu\rangle \langle \mu|O|\nu\rangle\langle \nu | \Psi_0\rangle
e^{i (E_\mu-E_\nu)t}.  
\label{ds}
\end{equation}
In the thermodynamic limit expectation values of \emph{local operators}
approach time independent values. These can be determined by the
recently proposed quench action method \cite{ce-13}. The latter
results in a particular set of particle and hole densities and a
corresponding ``representative eigenstate''
$|{\boldsymbol{\rho}_{sp}}\rangle$ such that for local operators ${\cal
  O}$
\be
\lim_{t\to\infty}\langle \Psi(t) | {\cal O}| \Psi(t) \rangle= 
\langle\boldsymbol{\rho}_{sp}|{\cal O}|{\boldsymbol{\rho}_{sp}}\rangle.
\label{NESS}
\ee
The state $|{\boldsymbol{\rho}_{sp}}\rangle$, which depends on the initial
state of the system $|\Psi(0)\rangle$, will be referred to as the
stationary state towards which the system evolves
in the sense of \fr{NESS}. In general the determination of the
representative eigenstate is a non-trivial  task and it has been
carried out in only a few cases
\cite{de_nardis_wouters,amsterdam,budapest,de_luca,bse-14,idw-15}. 
In the following we report the exact analytical solution of this
problem for the quench that we are considering. 

\textit{The post-quench stationary state.} As we already noted, the
state $|{\boldsymbol{\rho}_{sp}}\rangle$ is specified by sets of particle
and hole densities $\{\rho_n(\lambda)\}$, $\{\rho^{h}_n(\lambda)\}$.
It is useful to work with the dimensionless variable $x=\lambda/\bar{c}$
instead of the rapidity $\lambda$, and with a slight abuse of notation
we will keep the same symbol for functions of $\lambda$ and of $x$
when this does not generate confusion. We further define $\eta_{n}(x)=\rho^{h}_{n}(x)/\rho_{n}(x)$.
Our result for the densities describing the stationary state is
\bea
\eta_{1}(x)&=&\frac{x^2[1+4\tau+12\tau^2+(5+16\tau)x^2+4x^4]}{4\tau^2(1+x^2)},
\label{eta_1}\\
\eta_{n}(x)&=&\frac{\eta_{n-1}\left(x+\frac{i}{2}\right)\eta_{n-1}\left(x-\frac{i}{2}\right)}{1+\eta_{n-2}(x)}-1\ , \ n\geq 2, 
\label{relation1}\\
\rho_{n}(x)&=&\frac{\tau}{4\pi}\frac{\partial_{\tau}\eta^{-1}_{n}(x)}{1+\eta^{-1}_{n}(x)}\ ,
\label{relation2}
\eea
where we have defined $ \eta_0(x)\equiv 0$ and  $\tau=1/\gamma$. Note that a relation analogous to (\ref{relation2}) was also found in the quench to the repulsive Lieb-Liniger model \cite{de_nardis_wouters}. Given
$\eta_1(x)$ in (\ref{eta_1}), Eqs. (\ref{relation1}) and
(\ref{relation2}) provide all other densities specifying the stationary state. Indeed, using the relation
$\rho_n^h(x)=\eta_n(x)\rho_n(x)$ one can readily see that $\rho_n(x)$
and $\rho^h_n(x)$ are written as rational functions 
 (the actual expressions getting lengthier as $n$ increases). 

\begin{figure}
\includegraphics[width=0.48 \textwidth]{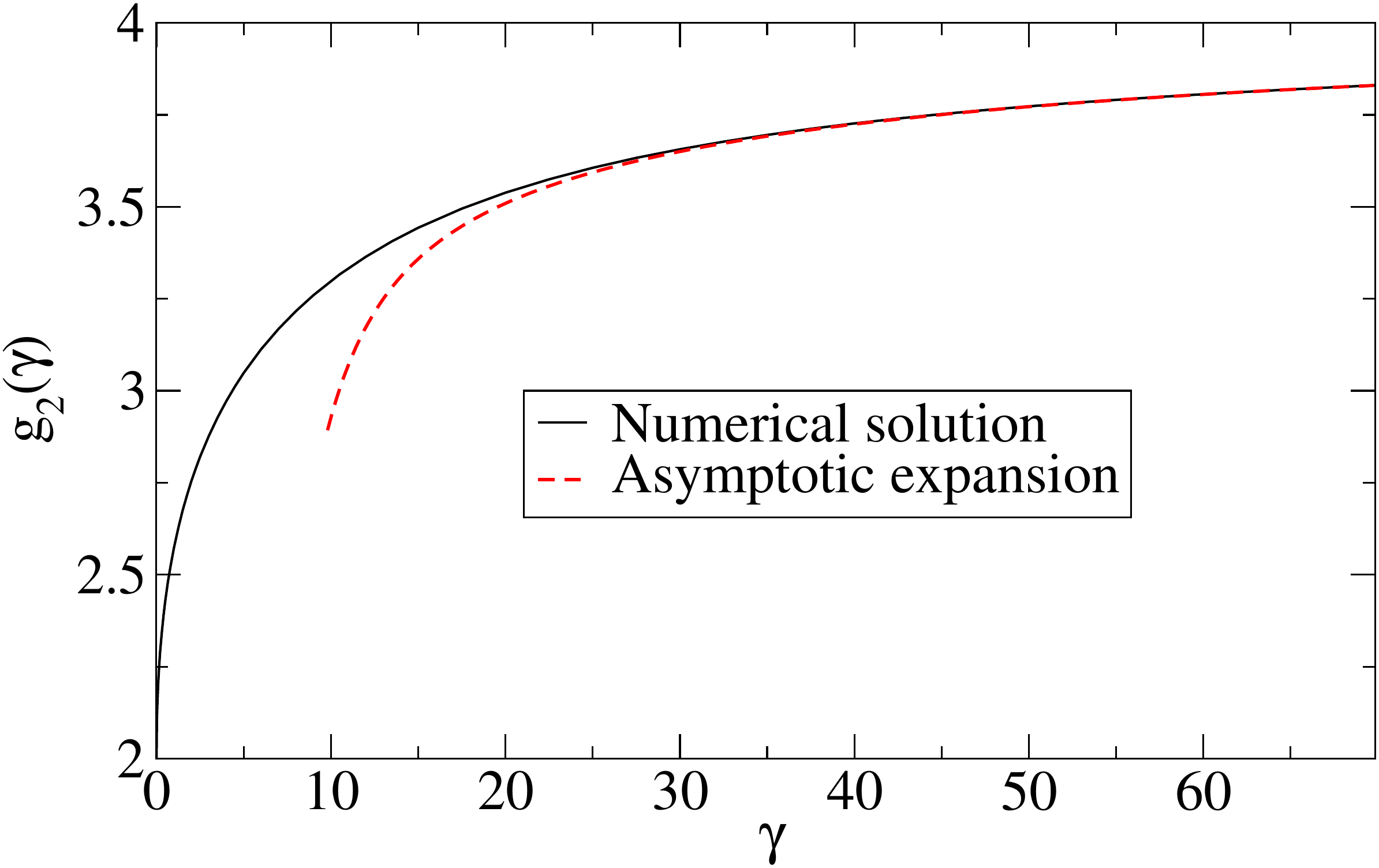}
\caption{(Color online) Numerical value of $g_{2}(\gamma)$ given in (\ref{one}) (solid line) and asymptotic analytical expansion as given in (\ref{g_2}) (dashed line).}
\label{fig_g2}
\end{figure}

The knowledge of the distributions $\rho_n(\lambda)$, $\rho^h_n(\lambda)$ allows in principle to compute the expectation value of all local observables in the post-quench stationary state. We considered the experimentally measurable local pair correlation (in the following equation $\hat \rho$ is the density operator)
\begin{equation}
g_2={\langle :\hat{\rho}^2(0):\rangle}/{D^2}\ .
\label{definition_g2}
\end{equation}
The computation of (\ref{definition_g2}) can be performed using the
Hellmann-Feynman theorem \cite{prep} and results in
\begin{multline}
  g_2=\gamma^2\sum_{m=1}^{\infty}\int_{-\infty}^{\infty}\frac{\mathrm{d}x}{2\pi}\ \left[2mxb_m(x)\frac{1}{1+\widetilde{\eta}_m(x)}
  \right. \\ \left.
  - 2\pi \widetilde{\rho}_m(x)\left(2mx^2-\frac{m(m^2-1)}{6}\right)\right]\ ,
\label{one}
\end{multline}
where the functions $b_{n}(x)$ are determined by
\begin{equation}
b_n(x)=nx-\sum_{m=1}^{\infty}\int_{-\infty}^{\infty}\mathrm{d}y\ \frac{1}{1+\widetilde{\eta}_{m}(y)}b_{m}(y)\widetilde{a}_{nm}(x-y).
\label{two}
\end{equation}
Here we defined $\widetilde{\eta}_{n}(x)=\eta_{n}(x\overline{c})$, $\widetilde{\rho}_{n}(x)=\rho_{n}(x\overline{c})$ (note that the computation of $g_2$ in (\ref{one}) requires the knowledge of the distributions $\eta_n$, $\rho_n$ characterizing the post-quench stationary state). Finally in (\ref{two}) we used $\widetilde{a}_{nm}(x)=a_{nm}(x\overline{c})$, where 
\begin{multline}
  a_{nm}(\lambda)=(1-\delta_{nm})a_{|n-m|}(\lambda)+2a_{|n-m|+2}(\lambda)+
   \\
 \ldots +2a_{n+m-2}(\lambda)+a_{n+m}(\lambda) ,
\label{def_a}
\end{multline}
and $a_{n}(\lambda)=\frac{2}{\pi n\overline{c} }\frac{1}{1+\left(\frac{2\lambda}{n \overline{c}}\right)^2}$. Eqs. (\ref{one}) and (\ref{two}) can be analytically solved for large $\gamma$ obtaining an expansion of $g_2$ in $1/\gamma$ for $\gamma\to\infty$. Up to the third order it reads
\begin{equation}
g_{2}(\gamma)=4-\frac{40}{3\gamma}+\frac{344}{3\gamma^2}-\frac{2656}{3\gamma^3}+\mathcal{O}(\gamma^{-4})\ .
\label{g_2}
\end{equation}
For generic $\gamma$ the sets of equations (\ref{one}) and (\ref{two})
can be solved numerically, resulting in the curve shown in Fig. \ref{fig_g2}. 
The function $g_{2}(\gamma)$ displays two intriguing features. The first one is that it is discontinuous in $\gamma=0$. Indeed
\begin{equation}
\frac{\langle \mathrm{BEC} |:\hat{\rho}^2(0):|\mathrm{BEC}\rangle}{D^2}=1\neq 2 =\lim_{\gamma\to 0} g_{2}(\gamma)\ .
\label{limit_g2}
\end{equation}
The second one is that in the limit $\gamma\to\infty$ it tends to the
finite value $g_2(\infty)=4$, cf. (\ref{g_2}). This is in contrast with
all other known stable situations, where the value of $g_2$ for infinite
interaction is always vanishing. This is true, for example, in the
repulsive regime at equilibrium (at finite or zero temperature)
\cite{gangardt} and crucially in the attractive regime for the
super-Tonks-Girardeau case, where $g_2$ is also vanishing for infinite
interactions \cite{abcg-05, bbgo-05, kmt-11}. In the following, we
argue that both of these behaviours can be ascribed to multi-particle
bound state effects. 

\begin{figure}
\includegraphics[width=0.48 \textwidth]{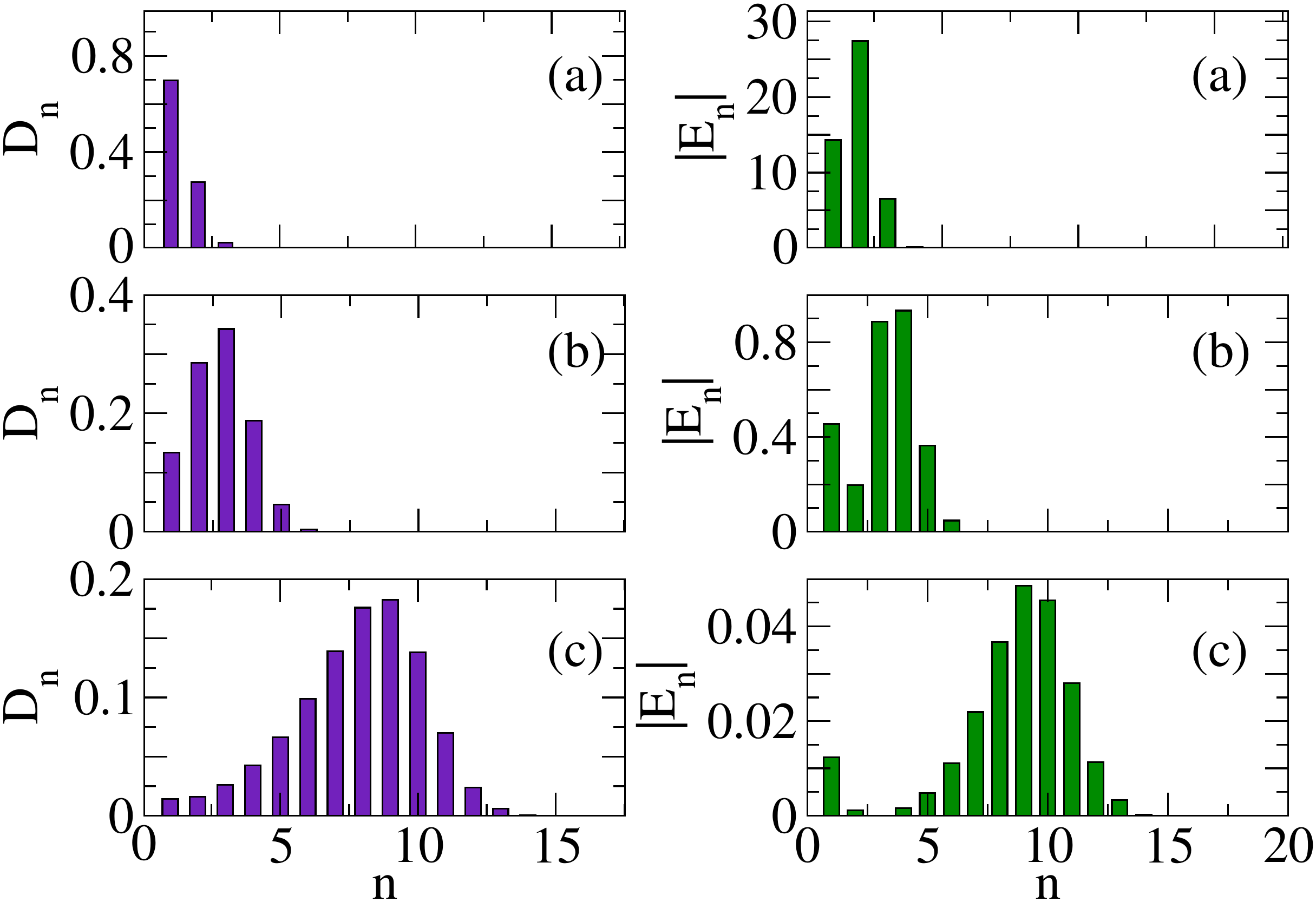}
\caption{(Color online) Density $D_n$ and absolute value of the energy $|E_n|$ of the particles forming $n$-strings, as defined in (\ref{d_n}), at fixed density $D=1$ and for (a) $\gamma=20$, (b) $\gamma=2$ and (c) $\gamma=0.2$.}
\label{isto}
\end{figure}

{\it Bound state content and physical implications}. The most
interesting property of our exact solution is that most of the
particles after the quench form bound states.
\begin{figure}
\centering
\includegraphics[width=0.48 \textwidth]{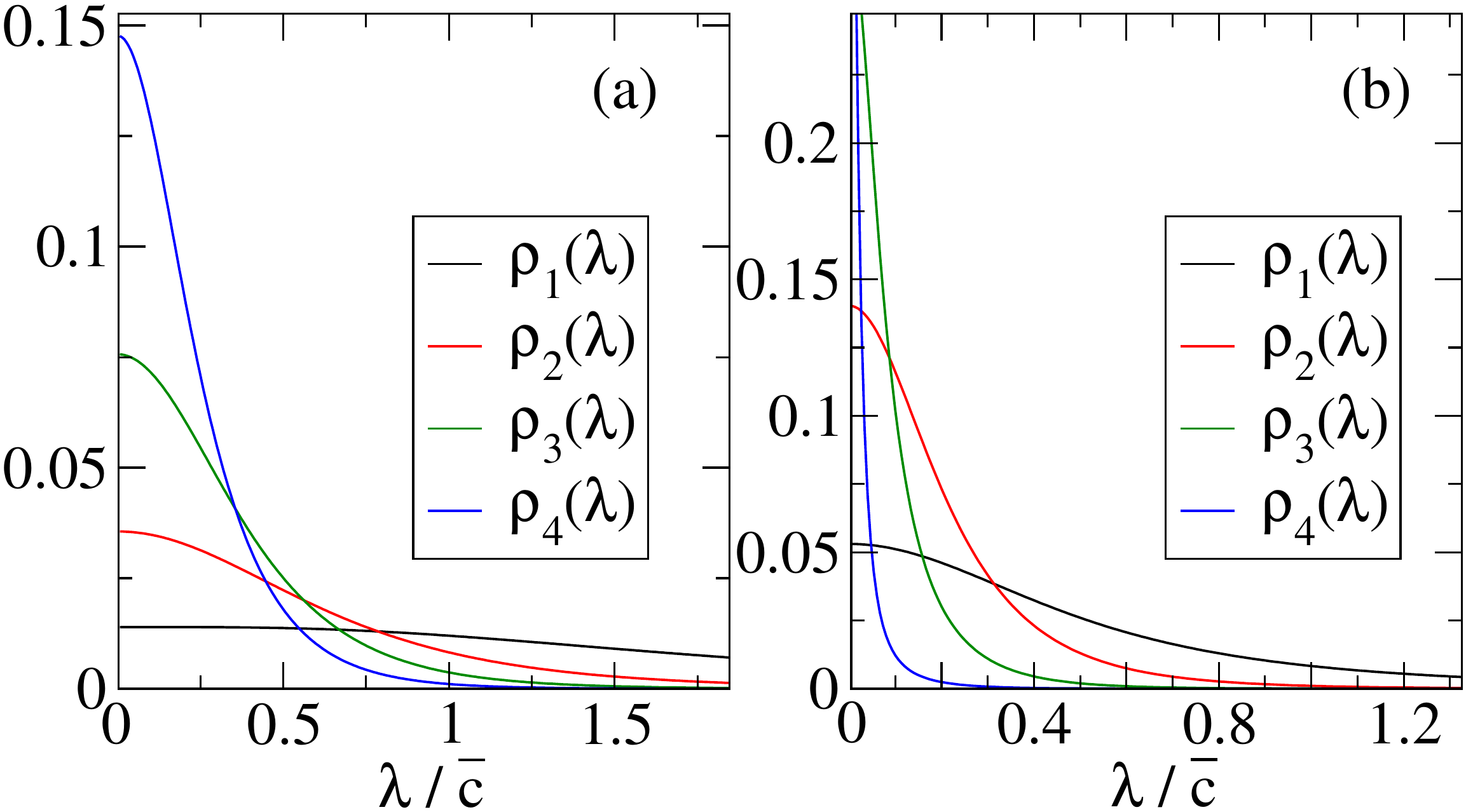}
\caption{(Color online) Density distributions for the string centers $\rho_{n}(\lambda)$ for (a) $\gamma=0.5$ and (b) $\gamma=2$.}
\label{densities}
\end{figure}
In Fig. \ref{densities} we display the particle densities for bound
states involving up to four particles for two values of $\gamma$.
We immediately see that bound particles outnumber unbound ones. To be
more quantitative, we define the density $D_n$ and the energy $E_n$ of
the particles forming $n$-strings as 
\be
D_n=n\int_{-\infty}^{\infty}\mathrm{d}\lambda\ \rho_{n}(\lambda),\quad
\label{d_n}
E_n=\int_{-\infty}^{\infty}\mathrm{d}\lambda\ \rho_{n}(\lambda)\varepsilon_n(\lambda) ,
\ee
where
$\varepsilon_{n}(\lambda)=n\lambda^2-{\overline{c}^2}n(n^2-1)/12$. In
terms of these quantities, the total density and energy are expressed
as sums of contributions arising from $n$-particle bound states
\begin{equation}
D=\sum_{n=1}^{\infty}D_n\ , \qquad \frac{E}{L}=\sum_{n=1}^{\infty}E_n\ . 
\end{equation}
In Fig. \ref{isto} $D_n$ and $|E_n|$ are plotted for decreasing value
of $\gamma=\overline{c}/D$ at a fixed density $D=1$. Fig. \ref{isto}
clearly shows that $n$-strings with $n\geq2$ in fact generally give
the dominant contributions. 

We now consider the dependence of the stationary state on $\gamma$ at
fixed density. We note that the total energy is conserved during the
quench and therefore is most easily calculated in the initial state
using Wick theorem, which gives
\begin{equation}
\frac{E}{L}=-\gamma D^3\ .
\label{total_energy}
\end{equation}
For large values of $\gamma$ the elementary bosons either remain
unbound, or form two-particle bound states, cf. Fig. \ref{isto}, while bound states
involving more than three bosons do not play an important role.
Moreover, it follows from (\ref{total_energy}) that $|E|$ is
large, which implies that the binding energy is very high and we are
dealing with tightly bound pairs of particles. These strongly affect
physical properties of the stationary state even in the limit
$\gamma\to \infty$. For example, the limiting value $g_2(\infty)$
can be imputed entirely to bound pairs \cite{prep}. As the value of
$\gamma$ decreases, bound states of increasingly higher numbers of
bosons become important. At the same time the magnitude of the total
energy $|E|$ is seen to decrease, and the binding energy eventually
approaches $0$ for $\gamma\to 0$. Thus, in the limit $\gamma\to 0$, 
heuristically,  the state after the quench is described by an
infinitely populated bound state having zero binding energy. We have
already alluded to the fact that $g_2(\gamma)$ exhibits a
discontinuity at $\gamma=0$, which has its origin in the presence of
multi-particle bound states for all positive $\gamma$. This is in
marked contrast to the situation found for quantum quenches in the
case of repulsive interactions \cite{de_nardis_wouters}. We note that $\gamma=0$ is a point of
non-analyticity for the solution of Eqs. (\ref{one}) and (\ref{two}),
making the limit $\gamma\to 0$ of $g_2(\gamma)$ in (\ref{limit_g2})
difficult to compute \cite{prep}.

\textit{Exact solution by the quench action approach}. We now sketch
the derivation of the results presented above. The basic idea is
to determine a representative eigenstate with the property that
expectation values of local operators in this state match their
stationary values reached at late times in the thermodynamic limit, cf.
\fr{NESS}. It was shown in Ref.~\cite{ce-13} that a state with this property
can be constructed as the saddle point of the so called quench 
action. The latter is given by
\be 
S_{QA}[\boldsymbol{\rho}]=2S[\boldsymbol{\rho}]-S_{YY}[\boldsymbol{\rho}],
\label{SQA}
\ee
where $\boldsymbol{\rho}$ generically indicates a state corresponding
to the sets of densities $\{\rho_n\}$ and $\{\rho_{n}^h\}$ and where 
$S_{YY}$ is the Yang-Yang entropy 
\be
\frac{S_{YY}[\boldsymbol{\rho}]}{L}= \frac{1}{2}\sum_{n=1}^\infty \int_{-\infty}^{\infty} d\lambda [\rho_n \ln (1+\eta_n)+ \rho_n^h \ln (1+\eta_n^{-1})]\ .
\label{yang}
\ee
The dependence of $S_{QA}$  on the initial state $|\Psi_0\rangle$
enters through $S[\boldsymbol{\rho}]=-\lim_{\rm th} {\rm
  Re} \ln \langle \Psi_0| \boldsymbol{\rho}\rangle$, where $\lim_{\rm th}$
denotes the thermodynamic limit $N,L\to\infty$ at fixed $D=N/L$.
In our case $|\Psi_0\rangle$ is the BEC state, which has vanishing
overlaps with non-parity invariant Bethe states (this is the reason of
the overall factor $1/2$ in (\ref{yang})) \cite{de_nardis_wouters}. The main 
ingredient in the quench action approach are the overlaps $\langle
\Psi_0| \boldsymbol{\rho}\rangle$ between the Bethe states and the
initial state. These are typically difficult to compute, but have been derived
for the Lieb-Liniger model in Refs.~\cite{de_nardis_wouters,brockmann_I}  
and applied to the attractive case in Ref. \cite{cd-14} (other
overlaps are also known \cite{p-13,pc-14,mazza-15}). In our case the functional
$S[\boldsymbol{\rho}]$ can be expressed as
\begin{equation}
S[\boldsymbol{\rho}]=\frac{LD}{2}\left(\ln\gamma+1\right)-\frac{L}{2}\sum_{n=1}^{\infty}\int_{0}^{\infty}d\lambda \rho_{n}(\lambda)\ln W_{n}(\lambda)\ ,
\end{equation}
where
\begin{equation}
W_n(\lambda)=\frac{1}{\frac{\lambda^2}{\overline{c}^2}\left(\frac{\lambda^2}{\overline{c}^2}+\frac{n^2}{4}\right)\prod_{j=1}^{n-1}\left(\frac{\lambda^2}{\overline{c}^2}+\frac{j^2}{4}\right)^2}\ .
\end{equation}
The saddle point conditions specifying the representative state
$|\boldsymbol{\rho}_{sp}\rangle$  are
\be
\left.\frac{\partial S_{QA}[\boldsymbol{\rho}]}{\partial \rho_n(\lambda)} \right|_{\boldsymbol{\rho}=\boldsymbol{\rho_{sp}}}=0, \qquad n\geq 1,
\label{oTBA1}
\ee
and take the form of coupled integral equations
\begin{equation}
  \ln\eta_{n}(\lambda)=-2hn-\ln W_{n}(\lambda)+\sum_{m=1}^{\infty}a_{nm}\ast\ln(1+\eta_{m}^{-1})(\lambda) ,
\label{coupled2}
\end{equation}
where we indicated the convolution between two functions with $f\ast g (\lambda)=\int_{-\infty}^{\infty}\mathrm{d}\mu f(\lambda-\mu)g(\mu)$ and where $a_{nm}(\lambda)$ is defined in (\ref{def_a}). Here the Lagrange multiplier $h$ has been introduced to fix the total
density $D$. The solution of Eq. (\ref{coupled2}) defines the
distributions $\eta_n(\lambda)$ for the saddle point state 
$|\boldsymbol{\rho}_{sp}\rangle$. A second set of equations is
obtained by taking the thermodynamic limit of the Bethe equations
(\ref{bethe_eq}), cf. \cite{cc-07,prep} 
\begin{equation}
\frac{n}{2\pi}-\sum_{m=1}^{\infty}\int_{-\infty}^{\infty}\mathrm{d}\lambda'a_{nm}(\lambda-\lambda')\rho_{m}(\lambda')=\rho_{n}(\lambda)(1+\eta_{n}(\lambda))\ .
\label{coupled}
\end{equation}
The sets \fr{coupled2}, \fr{coupled} of integral equations completely determine
the saddle point particle and hole densities. Their solution is given by
Eqs. (\ref{eta_1}), (\ref{relation1}) and (\ref{relation2}), with the relation $\tau=e^h$ \cite{prep}. 

{\it Experimental signature of the multi-particle bound states}.
Given that the quench from a BEC to a gas with attractive interactions
is clearly experimentally realizable as a simple modification of the
Super Tonks-Girardeau gas \cite{STG}, an obvious question is whether
there are smoking gun signatures of our novel state of matter. A key
feature of our steady state is the presence of several species of
bound states, which in the Lieb-Liniger model are infinitely long
lived as a consequence of integrability. It can be shown following
Ref.~\cite{bel-14} that different bound states have different group
velocities. This fact suggests that in the stationary state the
spreading of a local perturbation will exhibit several
``light-cones'', associated with different kinds of bound states
\cite{cc-06,cetal-12,fse-13,geiger-14}. We expect the situation to be 
analogous to what is seen theoretically \cite{gree12} and
experimentally \cite{fse-13} after local quantum quenches in the
Heisenberg XXZ chain in equilibrium. A detailed analysis of a local
quench in our steady state is however beyond the scope of this letter.
An important issue with regards to realizing our quench protocol in
cold atom experiments is the size of three-body losses. These can be
estimated from the three-body local correlation function $g_3$. The
calculation of $g_3$ in presence of bound states is a challenging
task, and we hope that our work will motivate studies in this direction. 

{\it Conclusions}. 
We have considered quantum quenches in the one dimensional Bose gas
with attractive delta-function interactions. In equilibrium
this model is known to be thermodynamically unstable, because it
supports the formation of many-particle bound states with binding
energies that scale like the third power of the number of atoms in the
bound state. The initial state in our quench is a BEC, and we consider
time evolution by the Lieb-Liniger Hamiltonian with arbitrary
attractive interaction strength. We have determined the exact steady state
reached at late times after this quench. This stationary state is
thermodynamically stable as a consequence of energy conservation.   
We have shown that this state is characterized by the presence of {\it
finite densities} of multi-particle bound states involving different
numbers of atoms. This composition, which automatically ensures
thermodynamic stability, distinguishes the stationary state in a clear
and {\it qualitative} way from other known stable states of the
model. In particular, our state differs significantly from the
super-Tonks-Girardeau gas, which is characterized by the {\it 
  absence} of bound states \cite{abcg-05, bbgo-05, mf-10,kmt-11}. 
A very interesting feature of our steady state is that the ``dominant''
species of multi-particle bound states can be changed by tuning
the value of the interaction strength, cf. Fig. \ref{isto}. We
have shown that the structure of the steady state results
in a value of $g_2$ between $2$ and $4$,
cf. Fig. \ref{fig_g2}, which is very different from other known
stable states in the Lieb-Liniger model. Finally we have argued
that these bound states can be
revealed by observing the spreading of local perturbations imposed at
late times after the initial quench.

{\it Acknowledgments}. 
This work was supported by the EPSRC under grants EP/I032487/1 and
EP/J014885/1 (FHLE),  the ERC under  Starting Grant
279391 EDEQS (PC and FHLE). FHLE thanks SISSA for hospitality.

\end{document}